\documentstyle[twoside,fleqn,espcrc2]{article}

\newcommand{\AmS}{{\protect\the\textfont2
\newcommand{\s}{\\ \vspace*{-3mm} }

  A\kern-.1667em\lower.5ex\hbox{M}\kern-.125emS}}

\newcommand{\lsim}{\raisebox{-0.13cm}{~\shortstack{$<$ \\[-0.07cm] $\sim$}}~}
\newcommand{\gsim}{\raisebox{-0.13cm}{~\shortstack{$>$ \\[-0.07cm] $\sim$}}~}

\newcommand{\s}{\\ \vspace*{-3mm} }

\newcommand{\non}{\nonumber}

\title{QCD Corrections in Supersymmetric Theories \hspace*{3cm} 
PM 97--33}

\author{Abdelhak Djouadi\address{Laboratoire de Physique Math\'ematique
        et Th\'eorique, UPRES--A 5032, \\
        Universit\'e de Montpellier II, F--34095 Montpellier Cedex 5, France}}
       
\begin{document}

\begin{abstract}
I discuss the effects of QCD radiative corrections in Supersymmetric theories. 
After summarizing the SUSY--QCD lagrangian in the Minimal Supersymmetric 
extension of the Standard Model, I will discuss the new features introduced 
by SUSY, and the main complications compared to standard QCD corrections. I 
will then discuss a few examples of QCD calculations in SUSY theories, for 
standard processes and for processes involving SUSY particles including the
extended Higgs sector. [Talk given at ``QCD 97", 
Montpellier 3-9 July 1997.]

\end{abstract}
\maketitle

\section{The SUSY--QCD Lagrangian}

Supersymmetry (SUSY) predicts the existence of a spin--zero partner for each 
Standard Model (SM) chiral fermion and a spin--1/2 partner for each gauge
boson or Higgs boson \cite{R1,R2}. Therefore, as strongly interacting 
particles one has in addition to gluons and quarks, the gluinos $\tilde{g}$ 
and three generations of left-- and right--handed squarks, $\tilde{q}_L$ and 
$\tilde{q}_R$. The interactions between gluon [$V^\mu$], gluino 
$[\lambda]$, quark [$\psi_i$] and scalar quark $[\phi_i$] fields are dictated 
by SU(3)$_{\rm C}$ gauge invariance and are given by the Lagrangian
\begin{equation}
{\cal L} = {\cal L}_{\rm kin} + {\cal L}_{\rm mat} + {\cal L}_{\rm self}+
{\cal L}_{\rm Yuk}+ {\cal L}_{\rm soft} 
\end{equation}
There is first the self--interactions of the gauge fields, where in addition
to the 3 and 4--gluon vertices that we do not write, there is a 
term containing the interaction of the gluinos with the gluons [$\sigma^\mu$ 
are the Pauli matrices which help to write down things in a two-component 
notation and $f_{abc}$ the structure constants of SU(3)]: 
\begin{eqnarray}
{\cal L}_{\rm kin}= ig f_{abc} \lambda^a \sigma^\mu \bar{\lambda}^b V_\mu^c
+``3V" + ``4V" 
\end{eqnarray}
Then there is a piece describing the interaction of the gauge and matter
particles
\begin{eqnarray}
&& {\cal L}_{\rm mat} = -g T^a_{ij}V_\mu^q \bar{\psi}_i \bar{\sigma}^\mu
\psi_j  \non \\
&& -ig T^a_{ij} V_\mu^q \phi_i^* \partial^{\leftrightarrow} \phi_j
+ g^2 (T^a T^b)_{ij} V^a_\mu V^{\mu b} \phi_i^* \phi_j  \non \\
&& +ig_Y \sqrt{2} T^a_{ij} (\lambda^a \psi_j \phi_i^*- \bar{\lambda}^a 
\bar{\psi}_i \phi_j) 
\end{eqnarray}
Besides the usual term  for the gluon--quark interaction and the terms 
for purely scalar QCD [the derivative term for the gluon--squark interaction
and the quartic term for the interaction between two gluons and two squarks] 
one also has a Yukawa--like term for the interaction of a quark, a squark and 
a gluino; SUSY imposes that the two coupling constants are the same $g_Y=g$. 

There is also a term for the self--interactions between the scalar fields; 
in the case where squarks have the same helicity and flavor, one has
\begin{eqnarray}
{\cal L}_{\rm self} = -g^2/3 \, ( \delta^{il} \delta^{kj} +
\delta^{ij} \delta^{kl}) \,  \phi_i \phi_j^* \phi_k \phi_l^*
\end{eqnarray}
Finally, there are the Yukawa interactions which generate the fermion 
masses, and the soft--SUSY breaking parameters which give masses to the gaugino
and scalar fields and introduce the trilinear couplings $A_q$. In the Minimal
Supersymmetric Standard Model (MSSM), where two Higgs doublets $H_1$ and $H_2$ 
are needed to break the electroweak symmetry \cite{R2}, these terms can be 
written in a simplified way for the first generation as [$u$ and $d$ are the 
left--handed quarks, $\tilde{u}$ and $\tilde{d}$ their partners and $Q/
\tilde{Q}$ the left--handed doublets]  
\begin{eqnarray}
{\cal L}_{\rm Yuk} &=& h_u Q H_1 u^c +h_d Q H_2 d^c \\
{\cal L}_{\rm soft} &=& - m_{\tilde{g}}/2 \,  \bar{\lambda} \lambda
+ \sum m_{\tilde{q}_i}^2 \phi_i^* \phi_i  + \cdots \non \\
&&+ h_u A_u \tilde{Q} H_1 \tilde{u}^c +h_d A_d \tilde{Q} H_2 \tilde{d}^c
+\cdots
\end{eqnarray}
Here some assumptions, such as R--parity conservation {\it etc...}, have 
been made; for a more detailed [and more rigorous] discussion see 
Ref.~\cite{R1}. 

We are now in a position to discuss the SUSY--QCD corrections to
physical processes.

\section{New features and complications compared to Standard QCD}

When one deals with calculations of QCD corrections in SUSY theories, 
a few complications compared to standard QCD corrections appear: 

-- Contrary to their standard partners the gluons, gluinos are massive 
particles due to the soft breaking of SUSY as discussed previously. 
In fact, gluinos are rather heavy in most of realistic and theoretically 
interesting models, and from the negative search of these states at the 
Tevatron a lower bound $m_{\tilde{g}} \gsim 150$ GeV has been set on their
masses \cite{PDG}. Light gluinos, which could be produced in 4--jet 
events at LEP1 seem to be experimentally ruled out \cite{zoltan}. Note 
also that gluinos are Majorana particles, and some care is needed in 
handling these states. 

-- The left-- and right--handed current eigenstates $\tilde{q}_L$ and 
$\tilde{q}_R$, mix to give the mass eigenstates 
$\tilde{q}_1$ and $\tilde{q}_2$ \cite{mixing}. The amount of mixing is 
proportional to the partner quark mass, and therefore is important only in 
the case of third generation, especially for the top squark. In 
fact for stops, the mixing can be so large that the lightest $\tilde{t}$ 
can be much lighter than the $t$ quark and all the other squarks. The mixing 
can also be important in the $\tilde{b}$ sector for large ${\rm tg}\beta$ 
[the ratio of the vev's of the two MSSM Higgs doublets] values. 

-- In standard QCD, the only parameters are the QCD coupling constant 
$\alpha_s$ as well as the quark masses $m_q$ which in the high--energy 
limit can be set to zero. In SUSY--QCD, much more parameters are present: 
besides the $\tilde{q}$ masses [which are different in general]  and the 
$\tilde{g}$ mass, one has the soft--SUSY breaking trilinear couplings $A_q$ 
as well as the mixing angles $\theta_q$. These parameters are in general 
related, complicating the renormalisation procedure and making 
next--to--leading order calculations more involved since one has to deal 
with loop diagrams involving different particles or with multi-particle 
final states with several different masses.  

-- There is also a problem with the regularisation scheme. Indeed, the usual 
dimensional regularisation scheme which is used in standard QCD, breaks 
Supersymmetry \cite{DRED}. 
For instance the equality between the strong gauge coupling $g$ and the 
Yukawa coupling $g_Y$ is not automatically maintained at higher orders, and one
has to enforce it by adding additional counterterms. In the dimensional 
reduction scheme, where only the four--vectors and not the Dirac algebra are 
in $n$--dimension, the equality between the two couplings is maintained 
automatically and this scheme is therefore more convenient. 
However, in some cases, gauge invariance can be broken in this scheme and 
again one has to add extra counterterms to satisfy the Ward identities.

-- Finally, there is an additional complication when Higgs bosons are involved. 
As already mentioned, at least two--Higgs doublets are needed in SUSY theories 
to break the electroweak symmetry, leading to the existence of at least five 
physical states: two CP--even $h/H$, one CP--odd $A$ and two-charged $H^\pm$
bosons. When calculating QCD corrections for the pseudoscalar Higgs 
boson $A$, one has to be careful with the treatment of $\gamma_5$ beyond the 
one--loop level.

\section{SUSY--QCD Corrections to Physical Processes}

\subsection{Standard Processes} 

I discuss now SUSY--QCD corrections to standard processes, i.e. processes
where only standard particles are involved in the initial and final state. 
Because no direct signal of SUSY particles has been observed directly, it is 
useful to look indirectly for SUSY in high--precision observables where 
SUSY loops effects can be important enough to alter the predictions of the SM. 
Of course, because of the large value of $\alpha_s$, the potentially largest 
effects are expected to come from corrections involving strong interactions.  

One of the simplest cases where SUSY--QCD corrections can be looked for is the 
cross section for $e^+ e^- \rightarrow {\rm hadrons}$. In addition to the 
standard corrections, virtual gluon exchange and gluon emission in the final 
state, one has also diagrams where squarks and gluinos are exchanged in the 
loops \cite{eehad}. Unfortunately, because gluinos and squarks are expected 
to have masses above 150 GeV, the corrections are rather small at present 
energies. For instance, for the hadronic width of the $Z$ boson, $\Gamma(Z 
\rightarrow \bar{q} q)$ , the SUSY--QCD correction is less than 0.2\% for 
realistic values of $m_{\tilde{g}}$ and $m_{\tilde{q}}$, which is less than the 
experimental accuracy of the measurement. Only for top quark
production at high--energy $e^+ e^-$ colliders, for instance at $\sqrt{s} 
\sim 500$ GeV, that the correction can reach the level of 1\% for
not too heavy squarks and gluinos. Thus, it will be very difficult to see 
any virtual effect from squarks and gluinos in $e^+ e^- \rightarrow q\bar{q}$,
except if these particles are light enough to be produced directly. 

Another measurable where strongly interacting SUSY particles can give large 
virtual effects is the $\rho$ parameter, defined as  the difference
between the $W$ and $Z$ boson self-energies at zero momentum transfer.
If there is a large splitting between the masses of the squarks which belong 
to the same weak isodoublet, for instance the $\tilde{t}/\tilde{b}$ doublet, 
the correction $\Delta \rho$ will grow with the square of the mass of the 
heaviest particle. This is similar to the SM case, where the $t/b$ doublet 
generates a correction which grows as $m_t^2$. $\Delta \rho$ enters all the 
high--precision measurements such as $\sin^2\theta_W$ or $M_W$, and one can 
constrain the masses of squarks by looking at the magnitude of their 
contributions. To make the constraints more precise, one needs to include QCD 
corrections. These two--loop corrections consist of diagrams with pure gluon 
exchange, pure scalar interactions and the gluino-quark--squark exchange 
diagrams, plus the corresponding counterterms. The calculation has been done 
recently \cite{deltarho} and it turns out that the QCD correction can be 
large, reaching the level of 30\%. They are in general positive, therefore 
increasing the sensitivity of electroweak observables to the contributions of 
squarks. 

Finally we have also the SUSY--QCD corrections to the top quark production 
at hadron colliders, $pp \rightarrow t\bar{t}$ \cite{ttbar}, and for the
top quark main decay mode, $t \rightarrow bW^+$ \cite{tdecay}, involving 
$\tilde{t}$-$\tilde{b}$--$\tilde{g}$ virtual exchange. For $t\bar{t}$ production
at the Tevatron, the corrections are of ${\cal O}(10\%)$ and therefore small;
at the LHC, the corrections can be larger but will be difficult to see due
the hostile environment at hadronic machines. For the top quark main decay 
mode, the SUSY--QCD corrections are even smaller, being at best a few percent. 
Therefore virtual effects of SUSY particles will be also difficult to isolate 
in this case.

\subsection{SUSY Processes}

Another aspect that I discuss now, 
is the QCD corrections to processes involving SUSY particles
in the final state. As usual, in order to have full control on the 
theoretical predictions for production cross sections and for decay widths, 
one needs to include the QCD corrections, which in general turn out to be 
rather large. 

The simplest process in this context, is the SUSY--QCD corrections to the 
production of scalar quark pairs in $e^+ e^-$ annihilation, $e^+ e^- 
\rightarrow \gamma / Z \rightarrow \tilde{q} \tilde{q}$. Part of the QCD 
corrections, the one due to pure gluon exchange and final gluon emission for 
equal mass squarks, can be 
adapted from Schwinger results for scalar QED \cite{schwinger}. But in SUSY 
theories, one needs to include first the gluino--quark exchange diagrams, and 
second to consider the case where the two squarks [in both the loops and 
the final state] have different masses. The calculation has been done by two 
independent groups \cite{eesqsq} and the results can be summarized as follows: 
for very large $m_{\tilde{g}}$, the gluino decouples and  one is left only
with the QED--like corrections; the corrections are of ${\cal O}(+15\%)$ for 
small $m_{\tilde{q}}$ [i.e. three times more than for quark pair production], 
and increase with increasing $m_{\tilde{q}}$; because of the Coulomb 
singularity, the 
correction blows up near threshold and non--perturbative effect must be 
included. For the gluino--exchange contribution, the correction is different 
for the partners of the light quarks and for the top squarks, 
because of the large value of $m_t$ and also because of the possible large
mixing in the stop sector; it is in general rather 
small and tend to decrease the cross section.

Another important process  where SUSY--QCD corrections are very important is 
the the production of squark and gluino pairs at hadron colliders,
$p\bar{p}/pp \rightarrow \tilde{q} \tilde{q}, \tilde{g}\tilde{g}, \tilde{q}
\tilde{g}$.  In this case, one has a large number of Feynman diagrams to 
consider, with loops involving gluinos, squarks and quarks making the 
calculation rather involved especially in the case of the $\tilde{t}$ squark 
where mixing should be included. The calculation has been recently completed
\cite{ppsqsq} and the results are as follows: the theoretical prediction 
for all processes and for both LHC and Tevatron energies, are nicely 
stabilized by including the NLO corrections. The K--factors, $K=\sigma_{\rm
NLO}/\sigma_{\rm LO}$, depend strongly on the considered process. For processes
with $\tilde{g}$ final states, $pp \rightarrow \tilde{g}\tilde{g}, \tilde{q}
\tilde{g}$, the corrections are large [up to 90\%] and positive, while
for squark pair production they are moderate [up to 30\%]. 
The corrections exhibit a sizeable dependence on the squark masses. Comparison 
of  the NLO cross sections with those used for experimental studies at the 
Tevatron reveal that the bounds on gluino and squark masses can be raised by 
+10 to +30 GeV; for LHC, the shift in mass due to the inclusion of NLO 
corrections can go up to 50 GeV. 
  
SUSY--QCD corrections to various scalar quark decays are also available.
For instance, QCD corrections to the decays of squarks [including
$\tilde{t}$ squarks which needs a special treatment due the mixing and the 
large value of $m_t$] into their  partners quarks and charginos or neutralinos,
$\tilde{q} \rightarrow q \chi^0 ,q' \chi^\pm$, have been calculated by several
groups \cite{sqdecays1,sqdecays2} and have been found to be rather important 
since they can reach the level of 30 to 40\%. The SUSY--QCD corrections to the 
decays of squarks to quarks and gluinos, $\tilde{q} \rightarrow \tilde{g}q$, 
can be even larger, while they are moderate for the reverse decay $\tilde{g} 
\rightarrow \tilde{q} q$ \cite{sqdecays2}. All these corrections are positive
and increase the decay widths.

Finally, there are also QCD corrections to the SUSY decays of the top quark,
$t \rightarrow \tilde{t}_1 \chi^0$ \cite{top2} with $\tilde{t}_1$ the 
lightest top squark and $\chi^0$ the invisible lightest neutralino. If
the decay is kinematically allowed, the QCD corrections  increase the 
decay width significantly. 

\subsection{The Higgs sector} 

QCD corrections in the MSSM Higgs sector are very important for neutral
Higgs boson production at the LHC, and lead to significant effects in
the decays of the Higgs bosons into quark or squark pairs, or the decays 
of top quarks and squarks into Higgs particles. 

The main production mechanism of the SM neutral Higgs boson at the LHC is
the gluon--gluon fusion process, $gg \rightarrow H^0$, which proceeds mainly
through virtual  top quark loops \cite{glashow}. The two--loop QCD correction 
leads to a large K--factor, of about $K \sim 1.6$--$1.8$, almost independent 
of the Higgs mass and stabilizes the theoretical prediction nicely \cite{Kfac1}.
In the MSSM, two  additional points
have to be considered for the production of the neutral Higgs particles in
the gluon fusion mechanism.  

First, in the standard QCD corrections, one has to 
include the contribution of the $b$--quark whose couplings can be strongly 
enhanced for large values of tg$\beta$; one also has to consider the 
case of the pseudoscalar Higgs boson where subtle problems related to the
implementation of $\gamma_5$ will appear. The K--factors \cite{Kfac2}
also vary little with the Higgs boson mass in general, yet they are strongly 
dependent on tg$\beta$: for small tg$\beta$ their size is approximately
as in the SM, $K\sim 1.7$, but for large tg$\beta$ they are in general close 
to unity except for the $h$ boson when it is SM--like. 

In addition to the standard QCD corrections, one has to include the QCD 
corrections to the squark [mainly $\tilde{t}$] loops which could enhance 
the production cross section significantly for relatively light top
squarks, $m_{\tilde{t}} \lsim 350$ GeV. The K--factors in this case \cite{Kfac3}
are almost the same as for the quark contributions and therefore, one can
use the standard K--factor to correct the sum of the quark+squark contribution
at one--loop. Note that this result is obtained when the gluinos are very
heavy and decouple. 

These calculations can be applied to the reverse process, the decay of the 
Higgs bosons into two--gluons $h,H,A^0 \rightarrow gg$. The corrections are
very large, increasing the decay widths by approximately 70\%. The SUSY--QCD 
corrections to the decays of the five MSSM Higgs bosons into quark pairs
[the standard QCD corrections have to be supplemented by the gluino-squark
exchange contributions] have also been discussed; they turned out to be
rather significant, especially for large values of tg$\beta$; see 
Ref.~\cite{sola}. The QCD corrections to Higgs boson decays into scalar 
quarks have been also recently completed \cite{Hsquark}: they can be very 
large, altering the decay widths by an amount which can be larger than 
50\%; they are positive and strongly depend on the $\tilde{g}$ mass. 

SUSY--QCD corrections to various decay modes involving MSSM Higgs 
bosons, such as $t \rightarrow H^+ b$ \cite{sola2} and $\tilde{t}_2 
\rightarrow \tilde{t}_1h$ or $\tilde{t}_1 A$ \cite{Hsquark}, have also 
been considered and found to be significant. 

Last but not least, the QCD corrections to the relations between the MSSM 
Higgs boson masses \cite{hempf} are also very important. These two--loop 
corrections, for instance,  decrease the maximal value of the lightest MSSM 
$h$ boson mass by an amount of the order of 10 to 20 GeV. 

\section{Summary}

A large theoretical effort has been made in the recent years for the 
calculation of QCD corrections in Supersymmetric theories. These
corrections turned out to be very important for the production 
and the decays of SUSY particles, including the extended Higgs sector. 
For processes involving only standard particles, unfortunately, the 
SUSY--loop effects are in most cases rather small, if squarks and gluinos 
are too heavy to be directly produced. More work will be still needed in 
the future on this subject.

\end{document}